\documentclass[aps,pra,twocolumn,showpacs,superscriptaddress]{revtex4}
\usepackage{graphicx}
\usepackage{amsmath}
\usepackage{amssymb,amsthm}
\usepackage{hyperref}
\usepackage[normalem]{ulem}

\newcommand{\eq}{\begin{eqnarray}} 
\newcommand{\en}{\end{eqnarray}}
\def\bra#1{\mathinner{\langle{#1}|}}
\def\ket#1{\mathinner{|{#1}\rangle}}
\newcommand{\braket}[2]{\langle #1|#2\rangle}

\begin{document}
\title{Composite-boson approach to molecular Bose-Einstein condensates in mixtures of ultracold Fermi gases}

\author{P. Alexander Bouvrie}
\affiliation{Centro Brasileiro de Pesquisas F\'isicas, Rua Dr. Xavier Sigaud 150, Rio de Janeiro, RJ 22290-180, Brazil}
% \affiliation{Instituto Carlos I de F\'isica Te\'orica y Computacional, Universidad de Granada, E-18071 Granada, Spain}

\author{Malte C. Tichy}
\affiliation{Department of Physics and Astronomy, University of Aarhus, DK-8000 Aarhus C, Denmark}

\author{Itzhak Roditi}
\affiliation{Centro Brasileiro de Pesquisas F\'isicas, Rua Dr. Xavier Sigaud 150, Rio de Janeiro, RJ 22290-180, Brazil}


\date{\today}

\begin{abstract}
We show that an Ansatz based on independent composite bosons  [Phys. Rep. {\bf 463}, 215-320 (2008)] accurately describes the condensate fraction of molecular Bose-Einstein condensates in ultracold Fermi gases. The entanglement between the fermionic constituents of a single Feshbach molecule then governs the many-particle statistics of the condensate, from the limit of strong interaction to close to unitarity. This result strengthens the role of entanglement as the indispensable driver of composite-boson-behavior. The condensate fraction of fermion pairs at zero temperature that we compute matches excellently previous results obtained by means of fixed-node diffusion Monte Carlo methods and the Bogoliubov depletion approximation. This paves the way towards the exploration of the BEC-BCS crossover physics in mixtures of cold Fermi gases with arbitrary number of fermion pairs, as well as the implementation of Hong-Ou-Mandel-like interference experiments proposed within coboson theory. 
\end{abstract}

\maketitle

\section{Introduction}

The experimental achievement of Bose-Einstein condensates (BEC) \cite{AndersonEnsherEtal1995,DavisEtal1995,Bradley1995} has opened a new and exciting era in the field of atomic and molecular physics \cite{BlochDalibardZwerger2008,Giorgini2008,ChinGrimmEtal2010}. This phenomenon is a consequence of the characteristic many-particle statistics of bosons, which is reflected by the commutation relations of their respective creation and annihilation operators. Atoms and molecules, however, are ultimately made of fermions, which, with a sufficiently strong attractive interaction, can be bounded to form structureless bosons. Feshbach resonances in mixtures of cold Fermi gases have provided a procedure to control the interaction between fermions of different species, and, thus, a way to observe the formation of BECs of molecules \cite{GreinerRegalJin2003,JochimBartensteinEtal2003,ZwierleinStanEtal2003,BourdelKhaykovichEtal2004,PartridgeEtal2005,BartensteinEtal2005} as well as the BEC-BSC (superfluidity) crossover \cite{BartensteinEtal2004,BourdelKhaykovichEtal2004,PartridgeEtal2005}. The many-particle statistics are governed by the interaction between fermions, {\it e.g.}, the fraction of fermion pairs that condense decreases with the magnetic field across Feshbach resonance, from the limit of strong attractive interaction (BEC) to the repulsive interaction regime (BCS) \cite{BourdelKhaykovichEtal2004,PartridgeEtal2005}. This many-interacting-particle-problem constitutes a theoretical challenge that has been addressed with mean field theories \cite{SalasnichManiniParola2005,PongLaw2007}. These approaches, however, overestimate the width of the momentum distribution of the Fermi gas and consequently the condensate fraction. To describe the BEC-BCS crossover theoretically, one has to resort to Monte Carlo simulations \cite{Giorgini2005}. In the strong-interaction regime, the collective statistics of molecular BEC can be well approximated by the Bogoliubov quantum depletion theory \cite{Bogoliubov1947}.

In the theory of cobosons (composite bosons)\cite{CombescotMatibetEtal2008}, the wavefunction of a fermion pair $\ket{\Psi}$ is represented in second quantization by application of the creation operator $\hat c^\dagger$ of the two-fermion composite on the vacuum, $\ket{\Psi} = \hat c^\dagger\ket{0}$,  and the successive application of this operator defines a Fock state of $N$ identical fermion pairs
\begin{equation}
\label{NCob}
\ket{N} = \frac{\left(\hat c^\dagger \right)^N}{\sqrt{ N! \chi_{N}} } \ket{0}.
\end{equation}
The {\it normalization factor} $\chi_{N}$ \cite{CombescotLeyronasEtal2003} reflects how, in accordance with the Pauli principle, the fermion pairs must distribute themselves over the available single-fermion states associated to the internal degrees of freedom of the state $\ket{\Psi}$. In this description, the interaction between fermion pairs can only come from fermion exchanges \cite{CombescotMatibet2010}; in other words, the physics of the many-particle system emerges from the state $\ket{\Psi}$ together with the fermion exchange interaction among the $N$ fermion pairs, and both ingredients are reflected in $\chi_N$. 

Despite the fact that Bose-Einstein condensates (i.e.~Fock states) of atoms and molecules constitute immediate candidates for an application of the coboson framework, this connection has received little attention{\it, e.g.} a general formalism for composite bosons at finite temperature was developed in Ref.~\cite{CombescotShiauEtal2011} and the condensate fraction for Gaussian states was calculated \cite{LeeThompsonEtall2014}. In the last decade, coboson theory has been extensively applied to phenomena such as excitons \cite{CombescotShiau2015}, which feature long-range particle interactions. Nevertheless, it has been shown recently that in systems with attractive (short-range) interaction between different fermion species, as the interaction induced by Feshbach resonances in cold-atom gases, the potential corresponds to a one-body operator in the coboson subspace \cite{CombescotShiauChang2016}. As a consequence, the ground state of ultracold interacting Fermi gases with two balanced species can be approximated by $\ket{N}$ in the dilute gas regime. One of the advantages of coboson theory with respect to approaches based on the BCS ansatz or the Bogoliubov depletion, which are only valid in the many-particle limit, is that it describes the ground state of unpolarized Fermi gases composed by an arbitrary number of fermion pairs $N$. Here, we show that the finite condensate fraction of fermion pairs at zero temperature in experiments of ultracold interacting Fermi gases, is a compositeness effect accurately described by the coboson theory. We verify, therefore, that the ansatz \eqref{NCob} of independent composite bosons can be applied fearlessly to molecular BECs.

We start completely from scratch in a reductionistic manner, and establish the state $\ket{\Psi}$ that describes two bound fermion on the side of a Feshbach resonance where the scattering length is positive. With that state in hand, we study the emergent many-particle statistics of molecular Bose-Einstein condensates. By using tools borrowed from quantum information \cite{Law2005} and a discretization method to decompose the state $\ket{\Psi}$ describing the fermion pair in the Schmidt form \cite{WangLawChu2005}, we compute the entanglement between the compounds of the molecule, and the closely related {\it normalization ratio} $\chi_{N+1}/\chi_N$, which governs the particle statistics of $N$ fermion pairs in the state $\ket{N}$. 

In particular, we show that the normalization ratio exhibits a universal behavior with the parameter $(k_F a)^{-1}$ for all $N$, which can be understood as the ratio between interparticle spacing and the scattering length ($a$), and which fixes the energy scale of the system ($k_F = \left(6\pi^2 n \right)^{1/3}$ is the Fermi wave number of a non-interacting gas with atom-pair density $n = N/V$). This universality allows us to fully characterize the statistics of the system by the entanglement between the molecular constituents. We obtain the condensate fraction of fermion pairs within the coboson theory and compare our results to Bogoliubov quantum depletion and fixed-node diffusion Monte Carlo simulations (FN DMC) \cite{Giorgini2005}. Our results match the established approaches  remarkably well in the strong-interaction regime, and deviate slightly near unitarity. This deviation is due to the strong-binding approximation that we use on the wavefunction. However, the finite condensate fraction at unitarity predicted by coboson theory \cite{CombescotShiauChang2016} indicates that this simple model reproduces essential physics of the BEC-BCS crossover. We also show that, while the normalization ratio preserves its universal character from few to many fermion pairs, the condensate fraction does not behaves as universal observable for configurations with just few fermion pairs.

\section{Feshbach molecule model}
\label{SecFeshbachModel}

As a test, we apply the model of Feshbach molecule of Refs.~\cite{ChengChin2005,WasakKrychEtal2014} to the ${}^6{\rm Li}_2$ broad resonance in order to compute the molecular ground state $\ket{\Psi}$.  It consists of two coupled channels, namely an open channel $o$ (background or entrance channel) and a closed channel $c$. In Feshbach magneto resonances, these channels can be identified as two different hyperfine or spin states of the constituent fermionic atoms of a molecule, which couple via Coulomb or exchange interactions \cite{ChinGrimmEtal2010}. The Hamiltonian of two interacting atoms of mass $m$ in a harmonic trap is given by
\begin{equation}
\label{Hamiltonian}
H = -\frac{\hbar^2}{2m} \left(\vec \nabla_1^2 + \vec \nabla_2^2\right) + \frac{m\omega^2}{2} (r_1^2+r_2^2) + \hat V_\text{int}(\vec r_1-\vec r_2),
\end{equation}
where $\omega$ is the trapping frequency and $\hat V_\text{int}$ the potential energy of the atom-atom interaction. Using the center-of-mass $\vec R = (\vec r_1 + \vec r_2)/2$ and relative $\vec r = \vec r_1 - \vec r_2$ coordinates, the Hamiltonian factorizes as $H=H_R + H_r$ and the state of the system becomes separable: $\ket{\Psi} = \ket{\psi_R} \ket{\psi_r}$. The ground-state solution of the center-of-mass Schr\"odinger equation $H_R\ket{\psi_R}=E_R\ket{\psi_R}$ is given by the isotropic harmonic oscillator energy $E_R=E_{\text{h.o.}} = 3 \hbar \omega/2$ and the Gaussian function $\psi_R(R) = (\sigma^{2} \pi)^{-3/4} e^{-\frac{R^2}{2 \sigma^2}}$, where $2\sigma = \sqrt{2\hbar/m\omega}$ is the radius that characterizes the size of the spherical trap such that 95\% of the wave function $\psi_R(R)$ is confined in the volume $V=(4/3)\pi (2\sigma)^3$. 

The interaction between atoms is described by a spherical well potential $\hat V_\text{int}(r)$ with a finite range for the interaction $r_0$ \cite{ChengChin2005}; within the range of interaction, $r=|\vec r|<r_0$, the attractive potential of the closed (open) channel is given by $-V_c$ ($-V_o$), and outside the interaction range by $\infty$ ($0$). The two atoms couple to a molecular bound state with effective binding energy $E_m$ and their wavefunction, for the relative motion, is the most-weakly-bound molecular state of $H_r$ with energy $E_r=-E_m<0$. Experimentally, an external magnetic field $B$ induces a Zeeman shift in the energy level of both channels, such that $E_m$ and, consequently, the effective atom-atom interaction can be tuned with $B$. 

\begin{figure}[ht]
\includegraphics[scale=.8]{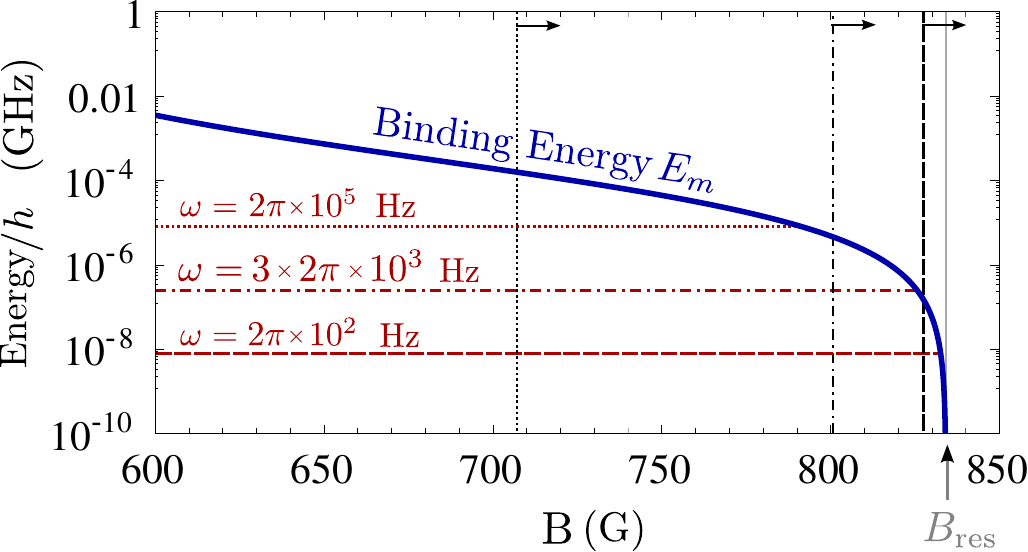}
\caption{Binding energy as a function of the magnetic field $B$ (solid line). Horizontal lines are the confining energies $E_\text{h.o.}/2$ for the frequencies $\omega = 2\pi \cdot 10^2$ Hz (dashed), $3 \cdot 2\pi \cdot 10^3$ Hz (dotted dashed) and $2\pi \cdot 10^5$ Hz (dotted). For these trapping frequencies, the BEC-BCS crossover extends to the strong-binding regime $(E_m>E_\text{h.o.}/2)$, i.e. from the vertical lines given by $k_F a = 1$ with $N=1$ to the right, as the arrows indicate. The unitary limit is found in the resonant position $B_\text{res}$ (vertical solid line).}
\label{FigureEnergy}
\end{figure}

To solve analytically the relative coordinate equation $H_r\ket{\psi_r}=E_r\ket{\psi_r}$ of this two-particle system, we assume that the binding energy is larger than the confining energy $2E_m>E_\text{h.o.}$. This approximation in the energy scale $2E_m>E_{ho}$ is equivalent to $a<2\sigma$ in the length scale. Experiments with Bose-Einstein condensates of Feshbach molecules are characterized by trapping frequencies $\omega$ of the order of $(2\pi)10^2$-$10^4$ Hz \cite{JochimBartensteinEtal2003,ZwierleinStanEtal2003}. In this frequency range the binding energy fulfils $2E_m>E_{\text{h.o.}}$ for a wide range of the magnetic field $B$, which extends to the BEC-BCS crossover region ($k_F a > 1$, where $k_Fa= (16/9\pi)^{-1/3}(\hbar/2m)^{1/2} a/\sqrt{\omega}$ for a single fermion pair $N=1$), see Fig.~\ref{FigureEnergy}. The details of the eigenenergy $(-E_m)$ equation and the eigenstate state $\ket{\psi_r} = \psi_o(r) \ket{o} + \psi_c(r) \ket{c}$ of the relative coordinate are provided in Appendix \ref{AppA}, together with further details about the two-state model for the ${}^6{\rm Li}_2$ molecule and the approximations that we use. 

The scattering length near the magnetic field location of the resonance, $B_{\rm res}$, is much larger than its off-resonance value $a_{\rm bg}$ ($a \gg |a_{\rm bg}|$) and can be approximated by \cite{ChengChin2005}
\begin{equation}
\label{ScatteringEqApp}
\frac{a-r_0}{a_{\rm bg}-r_0}= 1 + \frac{\Delta B}{B-B_{\rm res}}, 
\end{equation}
where $\Delta B$ and $B_{\rm res}$ are the resonance width and the resonance position, respectively, and $r_0$ (with $r_0\ll |a_{\rm bg}|$) is the interaction range. Although the actual ${}^6{\rm Li}_2$ Feshbach resonance has more than one closed channel and the potential energies of both closed $(V_c)$ and open $(V_o)$ channels are certainly not well potentials \cite{ChinGrimmEtal2010}, the simple model that we use reproduces essential physics of Feshbach resonances such as the binding energy or the channel mixing fraction \cite{ChengChin2005}. The binding energy of the molecule is $E_m \approx \hbar^2/[m(a-r_0)]^2$ for small $E_m$ and $a \gg r_0>0$.

\section{Schmidt decomposition and entanglement of a single molecule}

While every two-particle wavefunction admits a Schmidt decomposition, it is difficult -- if not impossible -- to find a close analytical solution for such a decomposition for wavefunctions in continuous variables  beyond the paradigmatic system of coupled oscillators \cite{Pruski1972}. Hence, we resort to the discretization method introduced in Ref.\cite{WangLawChu2005}, which exploits the cylindrical symmetry of a two-particle-systems by means of a Legendre expansion of the wavefunction. With this technique, the molecular wavefunction $\ket{\Psi}$ can be approximated by a Schmidt expansion 
\begin{eqnarray}
\label{WFSingleLabel}
\ket{\Psi} = \sum_{j}^S \sqrt{\lambda_{j}^o} \phi^{(o,1)}_{j}(\vec r_1) \phi^{(o,2)}_{j}(\vec r_2) \ket{o} + \nonumber \\
+\sum_{j}^S \sqrt{\lambda_{j}^c} \phi^{(c,1)}_{j}(\vec r_1) \phi^{(c,2)}_{j}(\vec r_2) \ket{c} ,
\end{eqnarray}
with finite Schmidt rank $S$. The Schmidt coefficients fulfill $0<\lambda_j^{o/c}<1$, and $\phi^{(o/c,1/2)}_{j}$ are the corresponding single-fermion states (normalized to unity) of atoms $1$ and $2$, in the open and closed channel, respectively. The Schmidt rank $S= (2l+1) \cdot (n_\text{max}+1) \cdot (l_\text{max}+1)$  is given by the number of coefficients $\lambda_j=\lambda_{nl}$ associated to the principal $n$ and angular momentum $l$  quantum numbers (with degeneracy $2l+1$) of the molecular state $\ket{\Psi}$ (see Appendix~\ref{AppB}). 

Close to unitarity, the Schmidt distribution of the closed channel becomes highly spread. In this limit, an accurate description of the wavefunction requires a very large number of Schmidt coefficients $\lambda_j^c$, making their calculation increasingly difficult. The closed channel contribution to the wave function $(\sum_{j=1}^S\lambda_j^c)$ is only relevant, however, in the weak-binding region $(k_Fa)^{-1}\ll 1$, near the resonant position \cite{ChengChin2005}. Therefore, we compute numerically only the Schmidt distribution of the open channel, since it is sufficient to obtain a minimal accuracy in the state normalization of $\bra{\Psi}\Psi\rangle\approx \sum_{j=1}^S \lambda_j^o>0.986$, up to the value $(k_Fa)^{-1}\approx 0.19$, with $n_\text{max}=350$ and $l_\text{max}=79$. For $(k_Fa)^{-1} < 0.19$ the closed channel contribution becomes relevant and the error in the state normalization increases drastically (see figure in the appendix). We will omit the index ``$o$'' in the open channel distribution $\lambda_{j}=\lambda_{j}^o$, which is the Schmidt coefficient distribution that characterizes the state $\ket{\Psi}$ hereafter. Note that, although the strong binding approximation, used in previous section~\ref{SecFeshbachModel}, restricts the validity of our results to the region  $(k_Fa)^{-1} > 0.5$, we extend the numerical simulation up to $(k_Fa)^{-1} = 0.19$.

\begin{figure}[ht]
\includegraphics[scale=0.72]{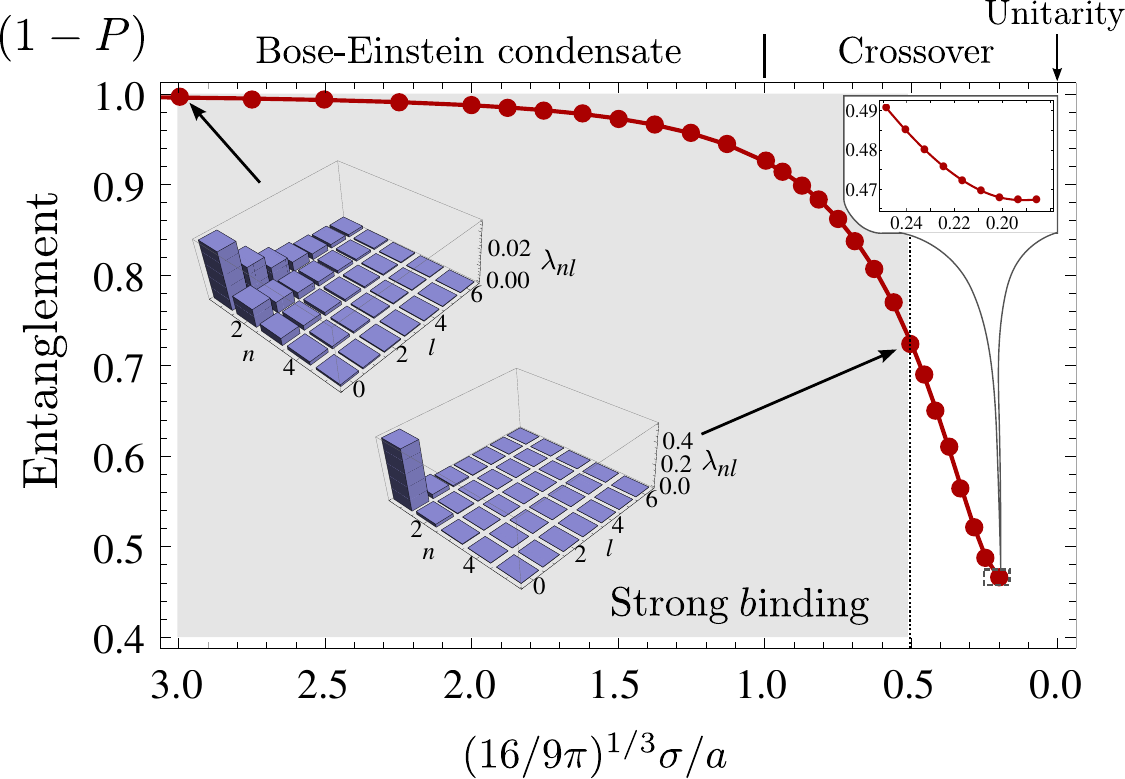}
\caption{Entanglement between the atoms comprising a Feshbach molecule (red dots, solid line guides the eye) as a function of the dimensionless parameter $(16/9\pi)^{1/3}\sigma/a =(k_Fa)^{-1}$ with $N=1$. The shaded area depicts the strong-binding region $2E_m>E_\text{h.o.}$ which is bounded by the vertical dotted line ($2E_m=E_\text{h.o.}$). The insets show the Schmidt coefficient distributions $\lambda_j=\lambda_{nl}$ of the state $\ket{\Psi}$ for $(k_Fa)^{-1}=3$ and $0.5$ (dots indicated by the arrows). The upper panel shows the finite entanglement in the limit of weak binding $k_Fa\gg1$ .}
\label{FigureEntanglement}
\end{figure}

With the above discretization method applied to the wavefunction $\ket{\Psi}$, we are able to quantify the entanglement between the fermionic atoms, along the positive scattering length region of the Feshbach resonance, {\it e.g.}, by means of $\mathcal{E} = 1-P$, the linear entropy, where $P = \sum_{j=1}^{S} \lambda_{j}^2$ is the purity of the single-fermion reduced density matrix. The entanglement of a single Feshbach molecule ($N=1$) depends only on the parameter $(16/9\pi)^{1/3}\sigma/a=(k_Fa)^{-1}$, since the Hamiltonian \eqref{Hamiltonian} can be rescaled by $\omega$ and written in terms of the ratio between the atom-atom interaction strength and the trapping frequency. When the magnetic field is ramped up, the scattering length of the ${}^6{\rm Li}_2$ Feshbach resonance increases. Consequently, the binding energy and entanglement decrease as shown in Fig.~\ref{FigureEntanglement}. In the BEC regime, the fermion pair is maximally entangled for $k_Fa\ll1$, and this entanglement is significantly decreasing up to the BEC-BCS crossover border ($k_Fa=1$). In the crossover region, the entanglement decreases drastically to a finite value ($\mathcal{E}_\text{min} \approx 0.47$, see inset panel of Fig.~\ref{FigureEntanglement}) in the limit of weak binding ($k_Fa\gg1$). %This entanglement could be extracted by delocalizing the atoms into two different wave packets \cite{GneitingHornberger2010}. 

\section{Bose-Einstein condensate of molecules}

We now turn to a many-body description of molecular Bose-Einstein condensates using  coboson theory. 

\subsection{Composite boson ground-state}

The wavefunction \eqref{WFSingleLabel}, which describes the fermion pairs in the molecular ground state $\ket{\Psi}$, has naturally motivated the introduction of composite boson creation operator  \cite{Law2005}
$
\label{CreationOp}
\hat c^\dagger=\sum_{j=1}^S \sqrt{\lambda_j}  \hat a^\dagger_j \hat b^\dagger_j ,%=\sum_{j=1}^S \sqrt{\lambda_j}  \hat d^\dagger_j ,
$
where $\hat a^\dagger_j$ ($\hat b^\dagger_j$) creates a fermion $1$ ($2$) in the Schmidt mode $\phi^{(1)}_{j}$ ($\phi^{(2)}_{j}$). Thus, the action of the operator $\hat c^\dagger$ on the vacuum describes the ground state of a Feshbach molecule $\ket{\Psi}= \hat c^\dagger \ket{0}$.  The $N$-fermion-pair Fock state \eqref{NCob},
\begin{equation}
\label{NCobState}
\ket{N} = \frac{1}{\sqrt{N!\chi_N}} \sum_{j_1\neq j_2\neq \cdots \neq j_N}^{1\le j_m\le S} \prod_{k=1}^N \sqrt{\lambda_{j_k}} \hat a_{j_k}^\dagger b_{j_k}^\dagger \ket{0} ,
\end{equation}
then describes a Bose-Einstein condensate of molecules, where the composite boson normalization factor is given by the elementary symmetric polynomial \cite{Law2005,TichyBouvrie2012a,TichyBouvrie2014} $\chi_{N} = N! \sum^S_{p_1 < p_2 < \cdots < p_N} \lambda_{p_1} \lambda_{p_2} \cdots \lambda_{p_N}$. The available recursive formula for the normalization factor \cite{TichyBouvrie2012a,TichyBouvrie2014} facilitates its computation for large number of Schmidt coefficients. 

As a consequence of the Pauli exclusion principle, the application of the creation operator $\hat c^\dagger$ on the $N$-coboson-state yields a sub-normalized state  \cite{ChudzickiOkeEtal2010} 
\begin{equation}
\hat c^\dagger \ket{N} = \sqrt{\frac{\chi_{N+1} }{\chi_N }} \sqrt{N+1} \ket{N+1} .  \label{addp}
\end{equation}
When a fermion pair in the state $\ket{\Psi}$ is added to the state $\ket{N}$, the fermion pair is accommodated among the $S-N$ unoccupied Schmidt modes, which only occurs with probability $\sum_{i \notin \{ j_1, \dots, j_N  \} }  \lambda_i$ for each configuration $j_1,\cdots,j_N$, see Eq.~\eqref{NCobState}. Therefore, the probability to successfully add a coboson to the state $\ket{N}$ is given by sum over the possible configurations of the set $j_1,\cdots,j_N$ \cite{TichyBouvrie2014}
\begin{equation}
\frac 1 {\chi_N} \sum_{j_1 \neq j_2 \dots \neq j_N}^{1 \le j_m \le S} \prod_{k=1}^N  \lambda_{j_k}  \left[ \sum_{i \notin \{ j_1, \dots, j_N  \} }  \lambda_i \right] 
= \frac {\chi_{N+1}} {\chi_N}   ,
\end{equation}
{\it i.e.}, by the normalization ratio $\chi_{N+1}/\chi_N$. Fermion pairs in the state $\ket{N}$ are correlated among themselves due to the Pauli principle. Hence, despite the molecular BEC being created by the $N$ successive addition of identical fermion pairs in the molecular ground state $\ket{\Psi}$ and the resulting state $\ket{N}$ constitutes the ground state of the $N$ fermion-pairs \cite{CombescotShiauChang2016}, there is no guarantee to find each of the $N$ fermion pairs in the state $\ket{\Psi}$, that is, $\ket{N}\not \propto \ket{\Psi}^{\otimes N}$.

\subsection{Condensate fraction: Comparison among theories}

The effective fraction of fermion pairs that populate the single-molecule ground state $\ket{\Psi}$ is given by the expectation value \cite{LeeThompsonEtall2014}
\begin{equation}
\label{CondFrac}
\frac{\bra{N} \hat c^\dagger \hat c \ket{N}}{N} = \frac{1}{N} + \left(1-\frac{1}{N} \right) \frac{\chi_{N+1}}{\chi_N} \le 1.
\end{equation}
The above expectation value constitutes, in fact, the condensate fraction of pairs at temperature $T=0$. This non-ideal bosonic condensation results from the competition of the constituent fermions to occupy the single-fermion states, or Schmidt modes, of $\ket{\Psi}$, and such competition increases with $a$ (see the Schmidt coefficients distributions of the insets in Fig.~\ref{FigureEntanglement}, which become more peaked (less uniform) for increasing $a$). 

\begin{figure}[ht]
\includegraphics[scale=.7]{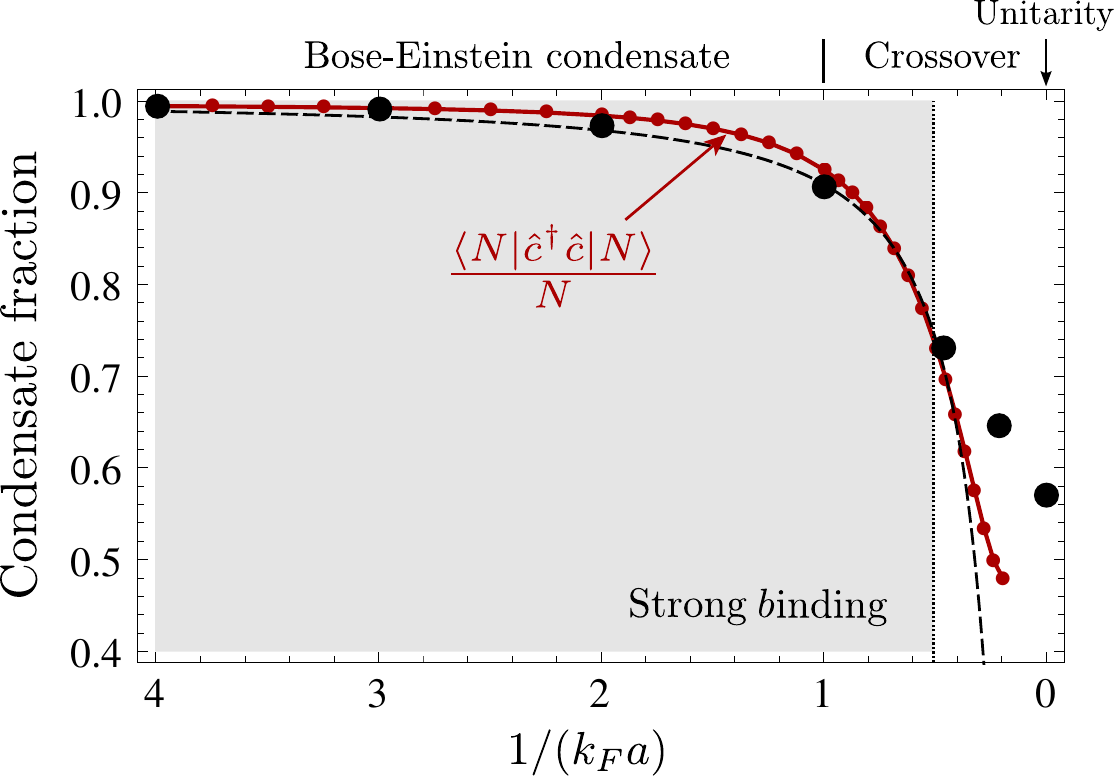}
\caption{Condensate fraction for $N=33$ predicted by coboson theory $\bra{N}\hat c^\dagger \hat c \ket{N}/N$ as a function of the dimensionless parameter $(k_Fa)^{-1}$ (red dots, solid line guides the eye). Black dots correspond to the FN DMC results \cite{Giorgini2005}, and the black dashed line to the Bogoliubov depletion of a Bose gas with $a_{dd}=0.6a$. The shaded area depicts the strong-binding region.}
\label{FigureCondensateFraction}
\end{figure}

In Fig.~\ref{FigureCondensateFraction} we show the effective fermion pair condensation across the positive scattering length resonance $a>0$, for $N=33$. The condensate fraction \eqref{CondFrac} matches qualitatively previous results obtained by FN DMC \cite{Giorgini2005}, and by the Bogoliubov quantum depletion approximation for a condensate of composite bosons $\alpha=1-8\sqrt{n a_{dd}^3}/(3\sqrt{\pi})$, where $a_{dd}=0.6 a$ is the characteristic dimer-dimer scattering length. While in the strong binding regime $(k_Fa)^{-1}>2$, the condensate fraction computed with coboson theory \eqref{CondFrac}(dots joined with solid line) is closer to the FN DMC simulations (big dots) than the Bogoliubov theory results (dashed line), in region $1<(k_Fa)^{-1}<2$, the Bogoliubov approach is closer to the FN DMC simulations than our results. All three results match excellently for $0.5<(k_Fa)^{-1}<1$. In the weak binding region, $(k_Fa)^{-1}<1$, the condensate fraction that we compute undermatches the FN DMC result due to the strong binding approximation that we use to compute the wavefunction (see Appendix~\ref{AppA}).

Nevertheless, coboson approach predicts finite entanglement in the limit $(k_Fa)^{-1}\to 0 $ \cite{CombescotShiauChang2016}, which leads to a finite condensate fraction at unitarity. The minimum condensate fraction close to unitarity that we computed, decreases with the number of fermion pairs $N$ as $(1+(N-1)\mathcal{E}_\text{min})/N$, in qualitative agreement with the FN DMC result \cite{Giorgini2005}. In this regard, the coboson theory reaches beyond mean-field and Bogoliubov theories, which makes it a candidate for the description of the BEC-BCS crossover.

\subsection{Universality of ultracold Fermi gases and the dilute regime}

The characteristic statistics of many-particle quantum systems, at the level of fermions and bosons, are established by the commutation relations satisfied by the respective creation and annihilation operators. Non-ideal bosonic operators $\hat c^\dagger$, however, obey non-conventional bosonic commutation relations, such that the expectation value of the commutator $[\hat c, \hat c^\dagger]$ on the state $\ket{N}$ reads \cite{Law2005,RamanathanKurzynski2011,Combescot2011}, 
\begin{equation}
\bra N  \left[ \hat c, \hat c^\dagger \right] \ket N = 2 \frac{\chi _{N+1}}{\chi _N} -1 , \label{commutatorexpl}
\end{equation}
When Eq.~(\ref{commutatorexpl}) equate to unity ($\chi_{N+1} /\chi_N=1$) the $N$ composite bosons behaves as ideal bosons, while deviations from unity entails observable consequences induced by the statistics of the constituent fermions \cite{CombescotBubinEtal2009,CombescotShiauChang2016,CombescotShiauEtal2011,KurzynskiRamanathan1012,TichyBouvrie2012b,Thilagam2013,LeeThompsonEtAl2013,LeeThompsonEtall2014}. The normalization ratio quantifies, therefore, the bosonic quality of the $N$ fermion pair state $\ket{N}$  \cite{TichyBouvrie2012a,TichyBouvrie2014}, and governs the many particle statistics of the system \eqref{CondFrac}. 

\begin{figure}[ht]
\includegraphics[scale=.7]{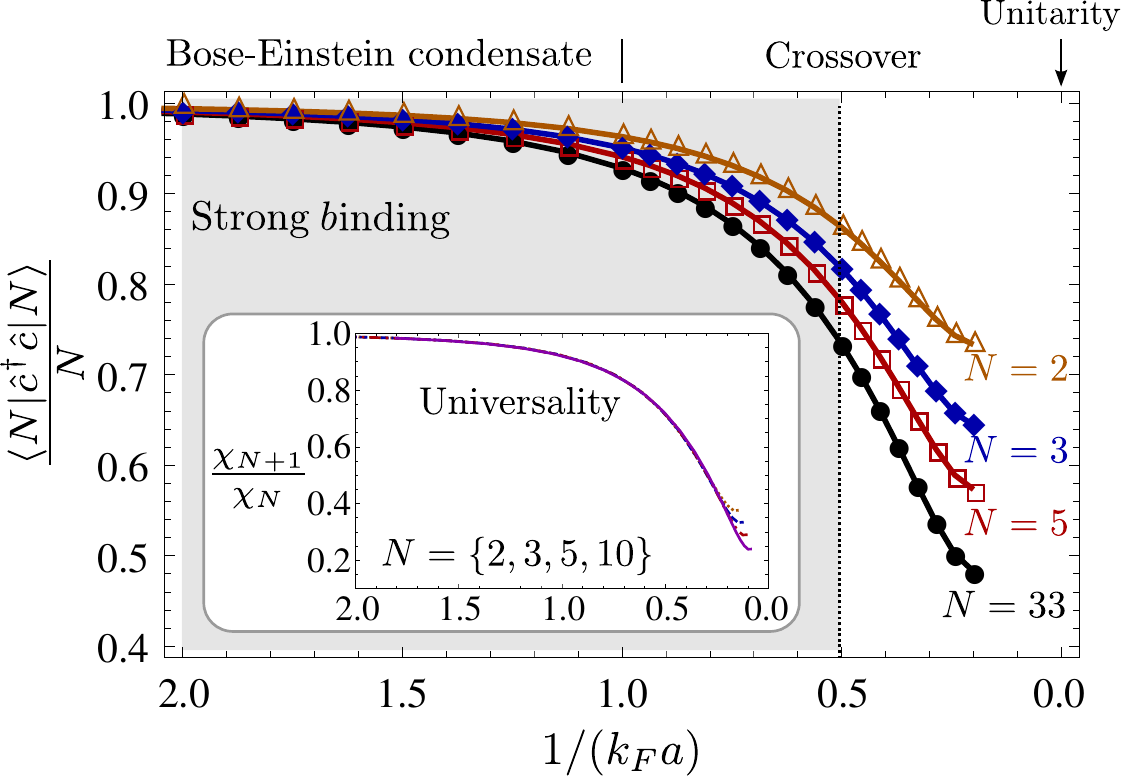}
\caption{Condensate fraction, $\bra{N}\hat c^\dagger \hat c \ket{N}/N$, as a function of $(k_Fa)^{-1}$ in the dilute regime with few fermion pairs, $N=2,3,5$ (yellow triangles, blue shaded squares and red squares, respectively), and with many fermion pairs $N=33$ (black dots). Different $N$ yield different curves, which reflects that the condensate fraction loses its universal character for few pairs. The inset shows the universality of the normalization ratio $\chi_{N+1} /\chi_{N}$ as a function of $(k_Fa)^{-1}$ for $N=2,3,5$ and $10$.}
\label{FigureCondensateFractionDilute}
\end{figure}

For interacting Fermi gases, several exact universal relations (Tan relations) have been shown, valid for all temperatures and spin compositions, which do not depend on details of the interparticle interaction.  The Tan relations connect a microscopic quantity, namely, the momentum distribution of the fermions, to macroscopic observables \cite{StewartEtal2010}. This universality of the macroscopic observables with the dimensionless parameter $(k_Fa)^{-1}$ is also reflected by the normalization ratio $\chi_{N+1} /\chi_N$. The inset of Fig.~\ref{FigureCondensateFractionDilute} clearly shows that the normalization ratio is a function only of the parameter $k_Fa$, {\it i.e.}, for a given $k_Fa$, the normalization ratio is independent of the number of fermion pairs $N$ such that all curves (for $N=2,3,5,10$) collapse, which also occurs for different trapping frequencies. We numerically find that $\chi_{N+1}/\chi_N$ evaluated at scattering length $a$ fulfills 
\begin{equation}
\left.\frac{\chi_{N+1}}{\chi_N} \right|_{a'=a} \approx \left.\frac{\chi_{N-m+1}}{\chi_{N-m}} \right|_{a'=\left(\frac{N}{N-m}\right)^{1/3}a} \approx \chi_2 |_{a'=N^{1/3}a}.
\end{equation} 
where $\chi_2=\mathcal{E}=1-P$. This relation significantly simplifies the computation of the normalization ratio for large $N$, which becomes computationally challenging otherwise \cite{ChudzickiOkeEtal2010,TichyBouvrie2012a,TichyBouvrie2014}.  The particle statistics of a molecular BEC is, therefore, fully characterized by the entanglement between the atoms that compose each single molecule, and can be controlled magnetically via Feshbach resonance. For instance, when manipulating magnetically the scattering length $a$ \eqref{ScatteringEqApp}, observables such as the condensate fraction of molecules depends uniquely on the entanglement of a single molecule, $\mathcal{E}|_{a'}$, evaluated at scattering length $a'=N^{1/3}a$, and the number molecules $N$. This reflects the capability of coboson theory to simplify the theoretical description of interacting cold Fermi gases.

While the normalization ratio of an $N$-coboson Fock-state preserves its universal properties with the parameter $(k_Fa)^{-1}$ for all $N>1$, other observables, such as the condensate fraction, lose their universal properties in the dilute regime of few fermion pairs. The dependence of such observables on the number of particles $N$ is not given only through the Fermi wave vector $k_F$. This is reflected in Eq.~\eqref{CondFrac} as a dependence of the condensate fraction on the normalization ration $\chi_{N+1}/\chi_N$ (and consequently on $k_Fa$) as well as on $N$. In Fig.~\ref{FigureCondensateFractionDilute} we show the condensate fraction as a function of $(k_Fa)^{-1}$ for few fermion pairs $N=2,3,5$. As a general trend, the larger the number of fermion pairs $N$ the lower the condensate fraction is for a given $k_Fa$, such that universality emerges only in the many particle limit ({\it e.g.}, $N=33$), where $\bra{N} \hat c^\dagger \hat c \ket{N}/N\approx \chi_{N+1}/\chi_N$.

\section{Conclusions and outlook}

In summary, the particle statistics of Bose-Einstein condensates of bi-atomic molecules is accurately described by a model of independent composite bosons. The condensate fraction of fermion-pairs in mixtures of Fermi gases is reproduced accurately in the strong-binding-regime at zero temperature. Within coboson theory, the state of one fermion pair completely characterizes  the collective statistics of the many-identical-pair-system due to fermionic exchange interaction. Thus, the entanglement between two fermions in a single-molecule bound state controls the many-particle statistics, since it approximately reflects  the ratio of fermions to available single-fermion states \cite{ChudzickiOkeEtal2010}, it leads to a universality of the normalization ratio with the dimensionless parameter $k_f a$ that characterizes the Fermi gas. Our result is valid for any species and number of bound fermion pairs, such that coboson theory allows to study ultracold interacting Fermi gases in the very dilute regime of few trapped fermion pairs \cite{SerwaneZurnEtal2011}; it predicts that some observables, such as the condensate fraction, lose their universal properties for few trapped fermion pairs, while other observables, such as the expectation value $\bra{N}\hat c^\dagger \hat c \ket{N}$, preserve their universal character. 

Because of the strong-binding approximation to the wavefunction, our theory deviates in the unitarity region, but the result of a finite value for the condensate fraction near unitarity indicates that the coboson theory is a strong candidate to explain the BEC-BCS crossover. In order to explore and demarcate the scope of coboson theory, it should be tested with more realistic wavefunctions, which could lead to an exact characterization of the BEC-BCS crossover up to unitarity, and also against observables \cite{StewartEtal2010} beyond the condensate fraction. Nevertheless, the present work underlines the fundamental importance of entanglement for the bosonic behavior of bound fermions in ultracold Fermi gases. The application of analytical bounds on the normalization ratio \cite{TichyBouvrie2012a,TichyBouvrie2014} will further simplify the evaluation of observables in molecular BECs.

When applied to atoms, coboson theory merely confirms that nucleus-electron-compounds constitute very good bosons indeed -- the purity of the single-electron density-matrix of a trapped hydrogen atom is of the order of $10^{-12}$ \cite{ChudzickiOkeEtal2010} -- rendering non-trivial predictions of deviations form ideal bosonic behavior or the observation of compositeness effects in atomic BECs infeasible. From that perspective, the application of coboson theory to the BEC-BCS-crossover is remarkable, as it provides a prominent system with experimentally accessible non-trivial compositeness effects, such that coboson theory yields a clear physical explanation for the finite condensate fraction: competition of the fermionic constituents for available single-fermion states. Consequently, coboson theory constitutes a powerful tool to explore theoretically new physical phenomena, and it can be applied fearlessly to molecular BECs. As a direct application, Hong-Ou-Mandel-like experiments \cite{HongOuMandel1987} of two interfering molecular BECs, in which the particle statistics are controlled by manipulating the inter-particle interaction, provide a wealth of observable effects of imperfect bosonic behavior \cite{TichyBouvrie2012b,BouvrieTichyMolmer2016}. Observables which manifest compositeness effects within coboson theory, such as second order correlation functions or the Mandel's parameter \cite{CombescotBubinEtal2009}, should be also tested experimentally. More in general, the present work underlines how the reductionistic coboson theory provides a physical explanation to why atoms, molecules and, in general, particles which are ultimately constituted by bound fermions, behave as bosons and are able to condense.

\section*{Acknowledgments}

We thank Luis D. Angulo and Tomasz Wasak for helpful comments on the numerical code and on the Feshbach model, respectively. We also thanks Rocio Jauregui, Klaus M\o{}lmer, Juan Omiste and Ana Majtey for stimulating discussions and the CSIRC of the University of Granada for providing the computing time in the Alhambra supercomputer.  P.A.B. gratefully acknowledges support by the Conselho Nacional de Desenvolvimento Cient\'ifico e Tecnol\'ogico (Brazilian agency) through a BJT {\it Ci\^encia sem Fronteiras} Fellowship, and by the Spanish MINECO project FIS2014-59311-P (co-financed by FEDER). I.R. also acknowledges partial support from CNPq.

\appendix

\section*{APPENDIX}

In this appendix we provide the details of the Feshbach molecule model presented in Refs.~\cite{ChengChin2005,WasakKrychEtal2014}, which we use in the main text to describe the molecular ground state $\ket{\Psi}$. The binding energy eigenequation and the scattering length are derived in the weak coupling regime (between the open and closed channels). The molecular wave function is computed assuming that the inter-atomic interaction energy is larger than the energy of the confining potential. The model is applied to the ${}^6{\rm Li}_2$ molecule. We also provide a description of the discretization method of Ref.~\cite{WangLawChu2005} which allows to obtain the approximate Schmidt decomposition of the molecular state.

\section{Feshbach molecule model}
\label{AppA}

Here we solve the system Hamiltonian (Eq.(2) of the main text) which describes two interacting particles in an harmonic trap
\eq
H = -\frac{\hbar^2}{2m} \left(\vec \nabla_1^2 + \vec \nabla_2^2\right) + \frac{m\omega^2}{2} (r_1^2+r_2^2) + \hat V_\text{int}(\vec r_1, \vec r_2),
\en
where $\omega$ is the frequency trapping, $m$ is the mass of the ${}^6{\rm Li}$ atom and $\hat V_\text{int}$ the potential energy of the atom-atom interaction. Using the center of mass, $\vec R = (\vec r_1 + \vec r_2)/2$, and the relative coordinate, $\vec r = \vec r_1 - \vec r_2$, the Hamiltonian factorizes as $H=H_R + H_r$ and the wave function of the system is separable: $\Psi(\vec r_1, \vec r_2) = \psi_R(\vec R) \psi_r(\vec r)$. The Schr\"odinger equation of the system $H\ket{\Psi} =E_\text{tot}\ket{\Psi}$ reads, therefore,
\eq
\label{SchCM}
\left[ -\frac{\hbar^2}{2m_R} \vec \nabla_R^2 + \frac{m_R\omega^2}{2} R^2 \right] \ket{\psi_R} = E_R \ket{\psi_R}  \\
\label{Schrel}
\left[ -\frac{\hbar^2}{2m_r} \vec \nabla_r^2 + \frac{m_r\omega^2}{2} r^2 + \hat V_\text{int}(r) \right] \ket{\psi_r} = E_r \ket{\psi_r}, 
\en
where $E_\text{tot} = E_R+E_r$, $m_R=2m$ and $m_r=m/2$. The ground state solution of the center of mass differential equation \eqref{SchCM} is given by the ground state eigenenergy of the isotropic harmonic oscillator $E_R=E_\text{h.o.}=3\hbar \omega/2$ and the Gaussian wave function
\eq
\psi_R(R) = \frac{1}{\sigma^{3/2} \pi^{3/4}} e^{-\frac{R^2}{2 \sigma^2}},
\en
where $2\sigma = \sqrt{2\hbar/m\omega}$ characterize the size of the harmonic trap.

We solve the Schr\"odinger equation of the relative coordinate \eqref{Schrel} in the strong binding regime, such that the scattering length is positive and lower than the harmonic oscillator length $0<a<2\sigma$. In this limit, the relative coordinate wave function is unaffected by the harmonic trap and the second term in the left side of Eq.~\eqref{Schrel} can be neglect. Thus, the relative coordinate Sch\"odinger equation can be approximated by
\eq
\label{SchrodingerEq}
E_r\ket{\psi_r} = \frac{\hbar^2}{m} (-\nabla_r^2+\hat V_\text{int}'(r) ) \ket{\psi_r},
\en
which has analytical solution \cite{ChengChin2005}. The fermion pair has a molecular bound state ($E_r=-E_m < 0$) near the continuum, and we assume that the binding energy of the atoms $E_m$ is larger than the confining energy of the molecule $2E_m>E_\text{ho}$.

The interaction potential $\hat V_\text{int} = \hbar^2/m \hat V_\text{int}'$ of the two-state model that we use to described a Feshbach molecule, is given by a spherical well potential with a finite interaction range $r_0$ \cite{ChengChin2005}:
\eq
\hat V_\text{int}' (r) = \left\{ 
\begin{array}{cc}
- \left(\begin{array}{cc}
  q_o^2 & \Omega \\
  \Omega & q_c^2 - \epsilon_c - \mu b
 \end{array}\right) & \text{For }r<r_0 \\
 ~ \left(\begin{array}{cc}
  0 & 0 \\
  0 & \infty
 \end{array}\right) & \text{For }r>r_0
\end{array}
\right. ,
\en
where the attractive potential of the closed (open) channel is given by $V_c=-\hbar^2 q_c^2/m$ ($V_o=-\hbar^2 q_o^2/m$) and $\Omega$ is the coupling amplitude between the closed and the open channels. The two atoms collide at energy $E_r=-E_m=-\hbar^2\epsilon_m/m$ in the entrance (open) channel and couple to the molecular bound state supported in the closed channel with an energy $E_c = \hbar^2 \epsilon_c /m$. Experimentally, an external magnetic field $B$ induces a Zeeman shift in the energy level of both channels, such that the energy between the continuum and the bare state can be tuned linearly with an energy given by $-\mu B=-\hbar^2 \mu b/m$, where $\mu=\mu_o - \mu_c$ and $\mu_o$ ($\mu_c$) is the magnetic moment of the atoms in the open (closed) channel. 

To solve the above coupled differential equation \eqref{SchrodingerEq} it is useful to introduce the superposition states 
\eq
\ket{+} &=& \cos \theta \ket{o} + \sin \theta \ket{c} \nonumber \\
\ket{-} &=& - \sin \theta \ket{o} + \cos \theta \ket{c} , 
\en
with $ \tan 2\theta=2\Omega /(q_o^2-q_c^2+\epsilon_c+\mu b)$. 
% Using the boundary conditions $\psi_c(r_0)=0$ and $\psi_o(r_0)/\psi'_o(r_0)=\psi_c(r_0)/\psi'_c(r_0)$ for the zero range scattering energy ($E=0$) solution 
% \eq
% \text{For~~} r>r_0: & \ket{\psi_r} \propto \frac{r-a}{r} \ket{o} ~~~~~~~~~~~~~~~~~~~~~~~~ \\ 
% \text{For~~} r<r_0: & \ket{\psi_r} \propto  \frac{\sin q_+ r}{r} \ket{+} + A \frac{\sin q_- r}{r} \ket{-},
% \en
% the following equation for scattering length $a$ is obtained \cite{ChengChin2005}:
The scattering length $a$ is defined from scattering in free space with zero relative kinetic energy. Hence solving Eq. \eqref{SchrodingerEq} for $E_r=0$ the following scattering length equation is obtained:
\eq
\label{ScatteringEq}
\frac{1}{r_0-a} = \frac{q_+ \cos^2\theta}{\tan q_+ r_0} + \frac{q_- \sin^2\theta}{\tan q_- r_0},
\en
where
\eq
q_+=\frac{\sqrt{q_c^2+q_o^2-\epsilon_c- \mu b + \left(q_c^2-q_o^2-\epsilon_c- \mu b\right) \sec 2\theta }}{\sqrt{2}} \nonumber \\
q_-=\frac{\sqrt{q_c^2+q_o^2-\epsilon_c- \mu b - \left(q_c^2-q_o^2-\epsilon_c- \mu b\right) \sec 2\theta }}{\sqrt{2}}
\en
are the eigen wave number associated with the states $\ket{+}$ and $\ket{-}$, respectively. 

Solving Eq. \eqref{SchrodingerEq} for a finite $E_m$, the relative coordinate wave function of the Feshbach molecule is given by 
\eq
\text{For~~} r>r_0: & \ket{\psi_r} = A_o \frac{e^{-\sqrt{\epsilon_m} r}}{r} \ket{o} ~~~~~~~~~~~~~~~~~~~~~~ \\ 
\text{For~~} r<r_0: & \ket{\psi_r} = A_+ \frac{\sin \bar q_+ r}{r} \ket{+} + A_- \frac{\sin \bar q_- r}{r} \ket{-},
\en
where $\bar q_\pm = (q_\pm^2-\epsilon_m)^{1/2}$. The constants $A_o$, $A_+$ and $A_-$ are obtained by means of the boundary conditions $\psi_c(r_0)=0$, $\psi_o(r_0^<)=\psi_o(r_0^>)$ and the normalization-to-unity $\braket{\psi_r}{\psi_r}=1$. The energy eigenvalue equation 
\eq
\label{EnergyEq}
-\sqrt{\epsilon_m}=\frac{\bar q_+ \cos^2\theta}{\tan \bar q_+ r_0} + \frac{\bar q_- \sin^2\theta}{\tan \bar q_- r_0},
\en
is determined by the condition $\psi_o(r_0)/\psi'_o(r_0)=\psi_c(r_0)/\psi'_c(r_0)$. 

In the weak coupling regime between the open and closed channels, it is legitimate to assume that $\Omega \ll q_o^2,q_c^2-\epsilon_c-\mu B$ and $|q_o^2-q_c^2+\epsilon_c+\mu B|$, such that, $\theta \ll1$, $q_+\approx q_o$ and $q_- \approx \sqrt{q_c^2-\epsilon_c-\mu B}$. In this limit, the closed channel contribution, which support the foreign bound state, is relevant only when it is close to the continuum and hence $\epsilon_c + \mu b\ll q_c/r_0$. In that case, the last term in Eq. \eqref{ScatteringEq} diverge, which implies that $\sin\sqrt{q_c^2-\epsilon_c-\mu b}\approx0$. Therefore, by performing the first order expansion on the last term of Eq. \eqref{ScatteringEq} and assuming that the middle term is roughly a constant $(r_0-a_{\rm bg})^{-1}$, the approximated scattering length, Eq. (3) in the main text
% \begin{equation}
% \label{ScatteringEqApp}
% \frac{a-r_0}{a_{bg}-r_0}= 1 + \frac{\Delta B}{B-B_{res}}, 
% \end{equation}
is obtained, where
\eq
\label{DelaB}
\Delta B = -\frac{\gamma \hbar^2}{\mu m} (a_{\rm bg}-r_0) \\
\label{Bres}
B_{\rm res}= - \mu^{-1} E_c +\Delta B ,
\en
are the resonance width and the resonance position, respectively, $a_{\rm bg}$ is the scattering background and $\gamma=2q_c^2\theta^2/r_0$. Since, in regime of bound states $E_r<0$ both channels are near the continuum, {\it i.e.} $|a_{\rm bg}|\gg r_0$ and $|\epsilon_c|\ll q_o/r_0$, performing the first order expansion on the last term of Eq.~\eqref{EnergyEq} leads to the eigenenergy equation 
\begin{equation}
\label{EigenenergyEq}
(\epsilon_m+\epsilon_c+\mu b)\left(\sqrt{\epsilon_m}-\frac{1}{a_{\rm bg}-r_0}\right) = \gamma,
\end{equation}
where $E_m=\hbar^2\epsilon_m/m$ is the binding energy.

Some parameters that characterize the model for the ${}^6{\rm Li}_2$ (of mass $m=6.02$ u) have been experimentally determined in Ref.~\cite{BartensteinEtal2005}:
$$
\begin{array}{|c|c|c|c|c|}
\hline
r_0 (a_0) & B_\text{res} (\text{G}) & \Delta B (\text{G})& a_{\rm bg} (a_0) & \mu (\mu_B) \\ \hline
29.9 & 834.15 & 300 & -1405 & 2.0 \\
\hline
\end{array}
$$
where $a_0$ is the Bohr radius, $\mu_B$ the Bohr magneton and $r_0$ is derived in \cite{GribakinFlambaum1993}. The remaining parameters are inferred by the model using the weak coupling approximation:
$$
\begin{array}{|c|c|c|c|c|}
\hline
\gamma^{-1/3} (a_0) & \epsilon_c (\hbar^2/m) & q_o^2 (\hbar^2/m) & q_c^2 (\hbar^2/m) & \theta \\ \hline
101 & -8.9\cdot 10^{17} & 9,688 \cdot 10^{17} & 6,308 \cdot 10^{19} & 0.0091 \\
\hline
\end{array}
$$
The parameter which characterizes the Feshbach strength $\gamma$ was obtained using Eq.~\eqref{DelaB}, the closed channel energy $E_c$ using \eqref{Bres}, and the eigen wave numbers $q_c$ and $q_o$ solving numerically the equations 
\eq
\frac{1}{r_0-a_{\rm bg}}=\frac{\sqrt{q_c-\epsilon_c} \cos^2\theta}{\tan \sqrt{q_c-\epsilon_c} r_0},
\en
and
\eq
\frac{1}{r_0-a_{\rm bg}}=\frac{q_o \cos^2\theta}{\tan q_o r_0},
\en
respectively, with $\theta=\sqrt{r_0  \gamma /2q_c^2}$.

\section{Schmidt decomposition}
\label{AppB}

Due to the cylindrical symmetry of the two-body system, the ground state wave function $\ket{\Psi}$ depends essentially on the radial coordinates of the atoms $r_1$ and $r_2$, and the inter-atomic angle $\gamma$. The center-of-mass and relative coordinates read, therefore, as $R=\sqrt{r_1^2+r_2^2+2r_1r_2\cos\gamma}/2$ and $r=\sqrt{r_1^2+r_2^2-2r_1r_2\cos\gamma}$, respectively. Thus, the wave function can be expanded in terms of the Legendre polynomial as
\eq
\Psi(\vec r_1,\vec r_2) = \Psi(r_1,r_2,\cos\gamma) = \sum_{l} \alpha_l(r_1,r_2) P_l(\cos\gamma), ~~~~~
\en
where the functions $\alpha_l(r_1,r_2)$ are given by
\eq
\alpha_l(r_1,r_2) = \frac{2l+1}{2} \int_0^\pi d \gamma \Psi(r_1,r_2,\cos\gamma) P_l(\cos\gamma) \sin\gamma. ~~~
\en

Using the spherical harmonics addition theorem $(2l+1) P_l(\cos\gamma) = 4\pi \sum_{m=-l}^l Y^*_{lm}(\theta_1,\varphi_1) Y_{lm}(\theta_2,\varphi_2)$, we have that 
\eq
\label{SemiSchmidtForm}
\Psi(\vec r_1,\vec r_2) = 4\pi \sum_{l} \frac{\alpha_l(r_1,r_2)}{2l+1} Y^*_{lm}(\theta_1,\varphi_1) Y_{lm}(\theta_2,\varphi_2).
\en
Note that this equation has the angular part already in the Schmidt form. To complete the Schmidt decomposition we have to perform the diagonalisation of the spatially discretized function $f_l(r_1,r_2)=r_1r_2\alpha_l(r_1,r_2)$ for each $l$, that is
\eq
f_l(r_1,r_2) = \sum_{n=0}^{n_\text{max}} k_{nl} u_{nl}(r_1) v_{nl}(r_2),
\en
where $k_{nl}$ are the eigenvalue of $f_l$, and $u_{nl}$ and $v_{nl}$ their corresponding eigenvectors. The prefactor $r_1r_2$ is necessary to ensure the correct normalization. That is, we need to diagonalize the matrix $M_{i,j}^{(l)} = \Delta r  f_l(\Delta r \cdot i, \Delta r \cdot j)$ with $\Delta r= r_{\rm max}/n_\text{max}$ and $i,j=1,2,\ldots,n_\text{max}$. The position $r_{\rm max}$ should be chosen as large as possible in the range were the wave function $\Psi(\vec r_1,\vec r_2)$ is mainly confined. This characteristic length of the relative coordinate $r_\text{max}$ depends on the trapping frequency $\omega$ and on the scattering length $a$. By using the $n_\text{max}$ eigenvalues, $k_{nl}$, and eigenvectors, $u_{nl}(r_1)$ and $v_{nl}(r_2)$, of $M_{i,j}^{(l)}$, together with Eq.\eqref{SemiSchmidtForm}, the Schmidt decomposition of the wave-function reads 
\eq
\label{SchimdtDecompWF}
\Psi(\vec r_1,\vec r_2) = \sum_{n=0}^{n_\text{max}} \sum_{l=0}^{l_\text{max}} \sum_{m=-l}^l \sqrt{\lambda_{nl}} \phi^{(1)}_{nlm}(\vec r_1) \phi^{(2)}_{nlm}(\vec r_2). ~~~~
\en
The $2l+1$ degenerated Schmidt coefficients are given by $\lambda_{nl} = 16\pi^2 k^2_{nl}/(2l+1)^2$, and the single fermion Schmidt eigenstates by $\phi^{(1)}_{nlm}(r_1)=\frac{u_{nl}(r_1) Y_{lm}(\theta_1,\varphi_1)}{r_1}$ and $\phi^{(2)}_{nlm}(r_2)=\frac{v_{nl}(r_2) Y_{lm}(\theta_2,\varphi_2)}{r_2}$, which form complete and orthonormal sets
\eq
\int d \vec r_1' \int d \vec r_2 \Psi(\vec r_1,\vec r_2) \Psi^*(\vec r_1',\vec r_2) \phi^{(1)}_{nlm}(\vec r_1') = \lambda_{nl} \phi^{(1)}_{nlm}(\vec r_1), \nonumber \\
\int d \vec r_2' \int d \vec r_1 \Psi(\vec r_1,\vec r_2) \Psi^*(\vec r_1,\vec r_2') \phi^{(2)}_{nlm}(\vec r_2') = \lambda_{nl} \phi^{(2)}_{nlm}(\vec r_2). \nonumber
\en

Since the open and closed channel states, $\ket{o}$ and $\ket{c}$, are orthogonal, the Schmidt decomposition \eqref{SchimdtDecompWF} applies separately to their respective wave functions $\psi_R(R)\psi_o(r)$ and $\psi_R(R)\psi_c(r)$. Therefore, the resulting Schmidt coefficient distribution, which describes state $\ket{\Psi}$, is given by the joint distribution $\{\lambda_{nl}^o,\lambda_{nl}^c\}$. We also introduce a single global label $j$ to characterize the $S = (2l+1) \cdot (n_\text{max}+1) \cdot (l_\text{max}+1)$ terms in \eqref{SchimdtDecompWF} corresponding to the indexes $n$, $l$ and $m$, such that $\lambda_j^{o/c}=\lambda_{nl}^{o/c}$. Thus, the molecular state $\ket{\psi}$ can be approximated by Eq. (4) in main text.

\begin{figure}[ht]
\includegraphics[scale=.72]{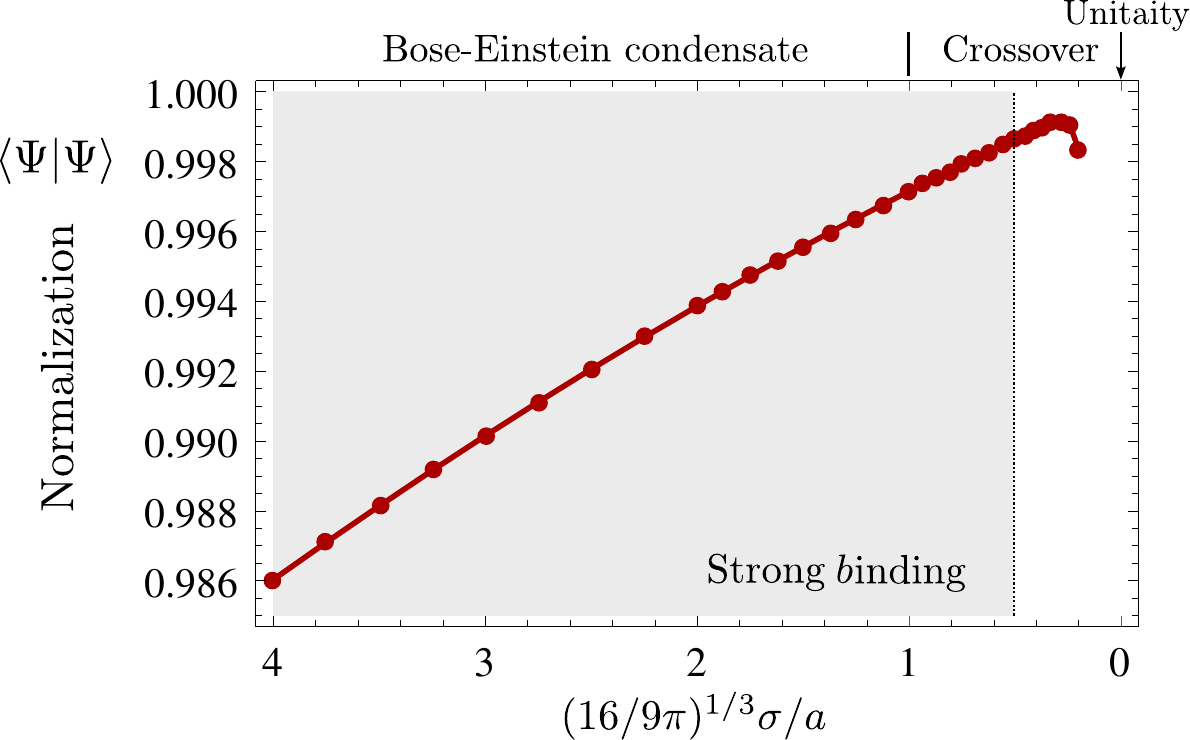} 
\caption[Fig.A]{Normalization of the wave function $\ket{\Psi}$ using the discretization method of Ref.~\cite{WangLawChu2005} with $n_\text{max}=350$, $l_\text{max}=79$, $r_\text{max} = 83338\cdot a_0$ and $\omega=2\pi \cdot 10^4$ Hz, as a function of $1/k_Fa=(16/9\pi)^{1/3}\sigma/a$ for $N=1$}
\label{NormalizationFig}
\end{figure}

The numerical computation of the Schmidt coefficient distribution $\{\lambda_1^o,\ldots,\lambda_S^o\}$ of the molecular ground state $\ket{\Psi}$ was performed with $\omega=2\pi \cdot 10^4$ Hz, $n_\text{max}=350$, $l_\text{max}=79$ and $r_\text{max} = 83338 \cdot a_0 \approx 6 \cdot \sigma$. Note that the characteristic length of the relative coordinate $r_\text{max}$ is larger than the of the center of mass coordinate length (or harmonic oscillator length)  $2\sigma$. The normalization of the wave function for these parameters is larger than $\bra{\Psi}\Psi\rangle\approx \sum_{j=1}^S \lambda_j^o>0.986$ in the range $4>1/k_Fa>0.19$, see Fig.~SM~\ref{NormalizationFig}. Close to the unitary limit, the contribution of the closed channel increases with a consequently abrupt decay of the state normalization.

\newpage

% \bibliography{Feshbach}

\begin{thebibliography}{10}

\bibitem{AndersonEnsherEtal1995}
M.~H. Anderson, J.~R. Ensher, M.~R. Matthews, C.~E. Wieman, and E.~A. Cornell,
\newblock Science {\bf 269}, 198 (1995).

\bibitem{DavisEtal1995}
K.~B. Davis {\em et~al.},
\newblock Phys. Rev. Lett. {\bf 75}, 3969 (1995).

\bibitem{Bradley1995}
C.~C. Bradley, C.~A. Sackett, J.~J. Tollett, and R.~G. Hulet,
\newblock Phys. Rev. Lett. {\bf 75}, 1687 (1995).

\bibitem{BlochDalibardZwerger2008}
I.~Bloch, J.~Dalibard, and W.~Zwerger,
\newblock Rev. Mod. Phys. {\bf 80}, 885 (2008).

\bibitem{Giorgini2008}
S.~Giorgini, L.~P. Pitaevskii, and S.~Stringari,
\newblock Rev. Mod. Phys. {\bf 80}, 1215 (2008).

\bibitem{ChinGrimmEtal2010}
C.~Chin, R.~Grimm, P.~Julienne, and E.~Tiesinga,
\newblock Rev. Mod. Phys. {\bf 82}, 1225 (2010).

\bibitem{GreinerRegalJin2003}
M.~Greiner, C.~A. Regal, and D.~S. Jin,
\newblock Nature {\bf 426}, 537 (2003).

\bibitem{JochimBartensteinEtal2003}
S.~Jochim {\em et~al.},
\newblock Science {\bf 302}, 2101 (2003).

\bibitem{ZwierleinStanEtal2003}
M.~W. Zwierlein {\em et~al.},
\newblock Phys. Rev. Lett. {\bf 91}, 250401 (2003).

\bibitem{BourdelKhaykovichEtal2004}
T.~Bourdel {\em et~al.},
\newblock Phys. Rev. Lett. {\bf 93}, 050401 (2004).

\bibitem{PartridgeEtal2005}
G.~B. Partridge, K.~E. Strecker, R.~I. Kamar, M.~W. Jack, and R.~G. Hulet,
\newblock Phys. Rev. Lett. {\bf 95}, 020404 (2005).

\bibitem{BartensteinEtal2005}
M.~Bartenstein {\em et~al.},
\newblock Phys. Rev. Lett. {\bf 94}, 103201 (2005).

\bibitem{BartensteinEtal2004}
M.~Bartenstein {\em et~al.},
\newblock Phys. Rev. Lett. {\bf 92}, 203201 (2004).

\bibitem{SalasnichManiniParola2005}
L.~Salasnich, N.~Manini, and A.~Parola,
\newblock Phys. Rev. A {\bf 72}, 023621 (2005).

\bibitem{PongLaw2007}
Y.~H. Pong and C.~K. Law,
\newblock Phys. Rev. A {\bf 75}, 043613 (2007).

\bibitem{Giorgini2005}
G.~E. Astrakharchik, J.~Boronat, J.~Casulleras, and S.~Giorgini,
\newblock Phys. Rev. Lett. {\bf 95}, 230405 (2005).

\bibitem{Bogoliubov1947}
N.~N. Bogoliubov,
\newblock J. Phys. USSR {\bf 11}, 23 (1947).

\bibitem{CombescotMatibetEtal2008}
M.~Combescot, O.~Betbeder-Matibet, and F.~Dubin,
\newblock Phys. Rep. {\bf 463}, 215 (2008).

\bibitem{CombescotLeyronasEtal2003}
M.~Combescot, X.~Leyronas, and C.~Tanguy,
\newblock Eur. Phys J. B {\bf 31}, 17 (2003).

\bibitem{CombescotMatibet2010}
M.~Combescot and O.~Betbeder-Matibet,
\newblock Phys. Rev. Lett. {\bf 104}, 206404 (2010).

\bibitem{CombescotShiauEtal2011}
M.~Combescot, S.-Y. Shiau, and Y.-C. Chang,
\newblock Phys. Rev. Lett. {\bf 106}, 206403 (2011).

\bibitem{LeeThompsonEtall2014}
S.-Y. Lee, J.~Thompson, S.~Raeisi, P.~Kurzyński, and D.~Kaszlikowski,
\newblock New J. Phys. {\bf 17}, 113015 (2015).

\bibitem{CombescotShiau2015}
M.~Combescot and S.-Y. Shiau,
\newblock {\em Excitons and Cooper Pairs: Two Composite Bosons in Many-Body
  Physics} (Oxford University Pess, 2015).

\bibitem{CombescotShiauChang2016}
M.~Combescot, S.-Y. Shiau, and Y.-C. Chang,
\newblock Phys. Rev. A {\bf 93}, 013624 (2016).

\bibitem{Law2005}
C.~K. Law,
\newblock Phys. Rev. A {\bf 71}, 034306 (2005).

\bibitem{WangLawChu2005}
J.~Wang, C.~K. Law, and M.-C. Chu,
\newblock Phys Rev. A , {\bf 72}, 022346 (2005).

\bibitem{ChengChin2005}
C.~Chin,
\newblock (2005),
\newblock arXiv:cond-mat/0506313.

\bibitem{WasakKrychEtal2014}
T.~Wasak {\em et~al.},
\newblock Phys. Rev. A {\bf 90}, 052719 (2014).

\bibitem{Pruski1972}
S.~Pruski, J.~Maćkowiak, and O.~Missuno,
\newblock Rep. Math. Phys. {\bf 3}, 227  (1972).

\bibitem{TichyBouvrie2012a}
M.~C. Tichy, P.~A. Bouvrie, and K.~M\o{}lmer,
\newblock Phys. Rev. A {\bf 86}, 042317 (2012).

\bibitem{TichyBouvrie2014}
M.~C. Tichy, P.~A. Bouvrie, and K.~M\o{}lmer,
\newblock App. Phys. B {\bf 117}, 785 (2014).

\bibitem{ChudzickiOkeEtal2010}
C.~Chudzicki, O.~Oke, and W.~K. Wootters,
\newblock Phys. Rev. Lett. {\bf 104}, 070402 (2010).

\bibitem{RamanathanKurzynski2011}
R.~Ramanathan, P.~Kurzy\'nski, T.~K. Chuan, M.~F. Santos, and D.~Kaszlikowski,
\newblock Phys. Rev. A {\bf 84}, 034304 (2011).

\bibitem{Combescot2011}
M.~Combescot,
\newblock Europhys. Lett. {\bf 96}, 60002 (2011).

\bibitem{CombescotBubinEtal2009}
M.~Combescot, F.~Dubin, and M.~A. Dupertuis,
\newblock Phys. Rev. A {\bf 80}, 013612 (2009).

\bibitem{KurzynskiRamanathan1012}
P.~Kurzy\'nski, R.~Ramanathan, A.~Soeda, T.~K. Chuan, and D.~Kaszlikowski,
\newblock New J. Phys. {\bf 14}, 093047 (2012).

\bibitem{TichyBouvrie2012b}
M.~C. Tichy, P.~A. Bouvrie, and K.~M\o{}lmer,
\newblock Phys. Rev. Lett. {\bf 109}, 260403 (2012).

\bibitem{Thilagam2013}
A.~Thilagam,
\newblock J. Math. Chem. {\bf 51}, 1897 (2013).

\bibitem{LeeThompsonEtAl2013}
S.-Y. Lee, J.~Thompson, P.~Kurzy\ifmmode~\acute{n}\else \'{n}\fi{}ski,
  A.~Soeda, and D.~Kaszlikowski,
\newblock Phys. Rev. A {\bf 88}, 063602 (2013).

\bibitem{StewartEtal2010}
J.~T. Stewart, J.~P. Gaebler, T.~E. Drake, and D.~S. Jin,
\newblock Phys. Rev. Lett. {\bf 104}, 235301 (2010).

\bibitem{HongOuMandel1987}
C.~K. Hong, Z.~Y. Ou, and L.~Mandel,
\newblock Phys. Rev. Lett. {\bf 59}, 2044 (1987).

\bibitem{BouvrieTichyMolmer2016}
P.~A. Bouvrie, M.~C. Tichy, and K.~Molmer,
\newblock Phys. Rev. A {\bf 94}, 053624 (2016)

\bibitem{GribakinFlambaum1993}
G.~F. Gribakin and V.~V. Flambaum,
\newblock Phys. Rev. A {\bf 48}, 546 (1993).

\bibitem{SerwaneZurnEtal2011}
F.~Serwane {\em et~al.},
\newblock Science {\bf 332}, 336 (2011).


\end{thebibliography}
% \bibliographystyle{h-physrev5}

\end{document}